# Introduction to Spectroscopy of Cr$^{4+}$:YAG Transparent Ceramics

Mykhailo Chaika[1,2]

[1]Institute of Low temperature and Structure and Structure Research, Polish Academy of Sciences, Okólna 2, 50 – 422 Wroclaw, Poland

[2]Graphene Energy Ltd, Curie-Skłodowskiej Str. 55/61, 50-369 Wrocław, Poland

**Abstract**

This paper focuses on the spectroscopic properties of Cr$^{4+}$:YAG transparent ceramic. Absorption, excitation, and emission spectra were measured over a temperature range from 5K to 300K. Low-temperature absorption spectra reveal sharp and narrow lines corresponding to partially allowed transitions from the ground state to the crystal field splitting components of the $^4T_2$ energy level. The shape of the excitation spectra was found to be independent of the monitored emission wavelength, indicating that Cr$^{4+}$ emission originates from the lowest excited state. Low temperature emission spectra exhibit a sharp and narrow ZPL, accompanied by the vibronic sidebands extending up to ~2000 cm$^{-1}$. Both absorption and emission spectra of the lowest excited state at low temperature consist of a doublet, with a splitting of 28 cm$^{-1}$. The temperature dependence of the spectroscopic parameters of this doublet is reported. Based on the obtained results, possible explanations of its origin are proposed.

*m.chaika@intibs.pl

## 1. Introduction

The application potential of Cr$^{4+}$:YAG laser material is not fully exploited due to the limited understanding of Cr$^{4+}$ spectroscopic properties. Over the last decades, Cr$^{4+}$-doped materials have found wide application in various fields, such as NIR LEDs, temperature sensors, and laser technologies [1,2]. Among these, the most significant is the development of tunable lasers operating in the 1200-1600 nm range, as well as passive Q-switched lasers operating at ~1 μm [3]. These lasers are typically based on tetrahedrally coordinated Cr$^{4+}$ ions in a garnet host, most commonly Y$_3$Al$_5$O$_{12}$ (YAG). The recently published "Cr$^{4+}$:YAG Crystals Market Outlook 2031 Report" highlights significant growth in the Cr$^{4+}$:YAG crystals market in recent years and projects continued expansion at least until 2031. However, further improvement of the laser performance of $Cr^{4+}$:YAG materials is limited by the lack of understanding of the nature of residual absorption of tetrahedral $Cr^{4+}$ ions. Despite the progress made over the last few decades, our understanding of the spectroscopic properties of $Cr^{4+}$ ions remains far from complete.

The laser performance of $Cr^{4+}$:YAG is attributed to the large absorption cross-section and relatively long lifetime of $Cr^{4+}$ ions. YAG has a cubic structure and belongs to the Ia-3d space group, with the stoichiometric formula $C_3A_2D_3O_{12}$, where C, A, and D denote dodecahedral, octahedral, and tetrahedral lattice sites, respectively. $Cr^{4+}$ ions can occupy both octahedral and tetrahedral sites. The possibility of using $Cr^{4+}$:YAG materials as Q-switched lasers is based on the large ground-state absorption cross-section and the relatively long excited-state lifetime of tetrahedrally coordinated $Cr^{4+}$ ions. It should be noted that $Cr^{4+}$ ions occupying octahedral sites do not exhibit luminescence. Therefore, to simplify the discussion in this paper, all references to the luminescence of $Cr^{4+}$:YAG will specifically pertain to $Cr^{4+}$ ions occupying a tetrahedral site.

The large absorption and lifetime enable $Cr^{4+}$ ions to saturate their absorption under laser radiation, allowing the development of pulsed lasers. The operation principle of passive Q-switched lasers is based on optical (phototropic) materials that change their transparency in response to absorbed energy [4]. These phototropic materials absorb the radiation emitted by the laser-active ions, thereby increasing the lasing threshold. Once the threshold is reached, a strong photon flux begins to build within the resonator, leading to the saturation of $Cr^{4+}$ ions' absorption. This rapidly reduces intracavity losses, and accumulated energy is released in a giant pulse. The phototropic absorber then begins to recover, preparing for the next cycle [5].

In the YAG lattice, $Cr^{4+}$ ions operate within a three-energy-level system, where photons are efficiently absorbed at allowed transitions with short lifetimes and subsequently stored in a lower-lying excited state with a longer lifetime. The main requirement for passive Q-switched materials is that the absorption cross-section of the saturable ions must be higher than the emission cross-section of the laser ions. Additionally, the upper energy level of the absorber should have a relatively long lifetime to deplete the ground state population by laser radiation. However, the fundamental challenge arises: large absorption cross-sections are typically associated with allowed transitions, which generally have short lifetimes [6]. In contrast, transitions with long lifetimes are often partially forbidden and thus exhibit low absorption cross-sections. To overcome this limitation, $Cr^{4+}$:YAG uses a three-level system, in which absorption occurs via an allowed transition, while an electrons are subsequently stored in a partially forbidden intermediate level (Fig. 1). Specifically, absorption of laser radiation take place thorough the allowed transition $^3B_1(^3A_2) \rightarrow {}^3A_2(^3T_1)$, denoted as transition (1)→(3). This is followed by a fast non-radiative transition from $^3A_2(^3T_1)$, which has a short lifetime, to the $^3B_2(^3T_2)$ energy level, labeled as transition (3)→(2). The $^3B_2(^3T_2)$ level has a relatively long lifetime, allowing electrons to become effectively "stored" at this energy level [7].

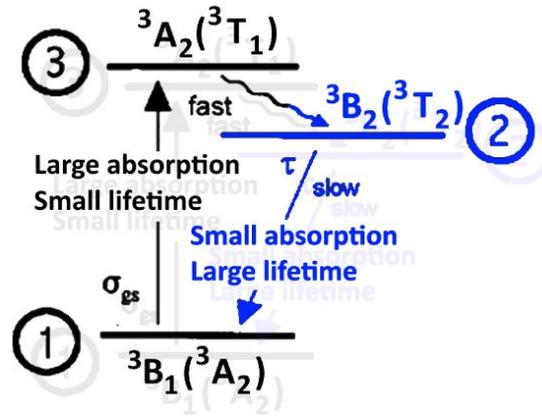

Fig. 1: Schematic illustration of the operating principle of a $Cr^{4+}$:YAG saturable absorber

The presence of multiple recombination channels from this "storage" level results in unsaturated losses that diminish the laser performance of $Cr^{4+}$:YAG. The laser properties of passive Q-switched lasers depend on several parameters, one of the most critical being the difference in absorption before and after saturation of the phototropic material [8]. One of the features of $Cr^{4+}$ ions is the presence of unsaturable losses. When a $Cr^{4+}$:YAG single crystal is irradiated with high-power 1064 nm laser light, the absorption due to $Cr^{4+}$ ions becomes saturated, with the exception of a residual absorption [9]. This residual absorption lead to a reduction in the overall laser performance of $Cr^{4+}$:YAG.

The unsaturated losses may be caused by the presence of differently oriented $Cr^{4+}$ centers, although other explanations may also be valid. Irradiation of the crystal with polarized light results in a twofold decrease in the residual absorption. Simultaneously, experimental measurements show that the laser performance of $Cr^{4+}$:YAG single crystal under polarizing pumping improves significantly. The slopes efficiency of laser generation increase from 8% under unpolarized light to 16% under polarized light, see Fig. 2 [10]. Kartazaev and Alfano suggested that this behavior is related to the low symmetry of the tetrahedral site in the YAG lattice occupied by $Cr^{4+}$ ions [10]. They proposed that the higher efficiency under polarized excitation arises from the selective excitation of different types of $Cr^{4+}$ oriented centers within the crystal [9]. To date, the origin of the observed residual unsaturable absorption in $Cr^{4+}$:YAG materials remains a gap in research.

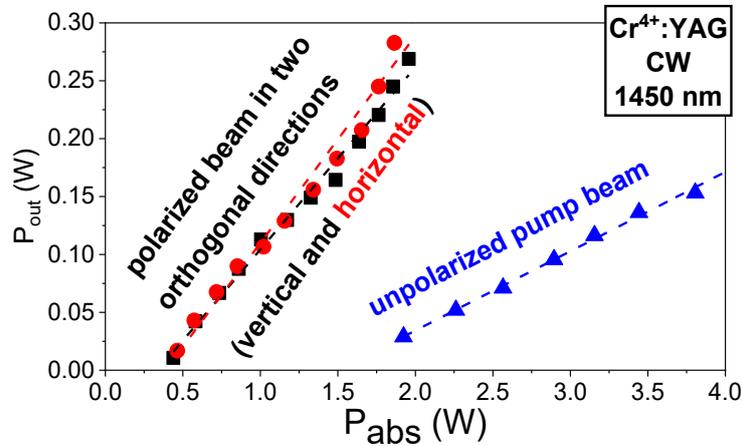

Fig. 2: $Cr^{4+}$:YAG laser output power at 1450 nm (reproduced from [10])

Excited state absorption, the presence of other absorbing centers, or various types of $Cr^{4+}$ centers have been proposed as possible origins of the unsaturated losses, but a satisfactory explanation requires further investigation. Several explanations have been proposed regarding the nature of residual absorption in $Cr^{4+}$:YAG materials. One of the most common hypotheses attributes this residual absorption to excited state absorption (ESA) of $Cr^{4+}$ ions. However, Okhrimchuk and Shestakov suggested that the unsaturable residual absorption does not originate from transitions within $Cr^{4+}$ centers themselves, but is instead associated with other, unidentified centers [9]. Hoffman et. all concluded that two types of $Cr^{4+}$ centers exist in $Cr^{4+}$:YAG, each characterized by slightly different sets of electron repulsion parameters [11]. It was found that the low-lying $^3B_2(^3T_2)$ emitting level (7875 cm$^{-1}$, or 1269 nm) is split into a doublet separated by 30 cm$^{-1}$, with each component exhibiting slightly different lifetimes. However, no further studies have been conducted to explore this phenomenon in detail.

The present paper focuses on the spectroscopic properties of $Cr^{4+}$ ions in $Cr^{4+}$:YAG transparent ceramic. The samples were synthesized using the Solid State Reaction (SSR) method, which involved vacuum sintering followed by air annealing. Excitation, emission, luminescence decay, and absorption spectra were recorded over a temperature range from 5K to 300 K. The present paper provides an overview of the key spectroscopic features in the absorption, emission, and excitation spectra. The influence of the temperature on the luminescence behavior is then discussed, with a particular focus on changes in the emission maximum, spectral width, and intensity of the narrow $Cr^{4+}$ lines. Finally, the results are interpreted in the context of the multisite nature of $Cr^{4+}$ emission centers in YAG ceramic.

## 2. Experimental

$Cr^{4+}$:YAG ceramics were obtained from CoorsTek research laboratory located in Uden, Netherlands. To carry out the optical investigation, the surface of the transparent Cr:YAG ceramics was polished with diamond abrasive. After processing, the pellets had a cylindrical shape with a diameter of 15 mm and a thickness of 3.8 mm.

X-ray diffraction (XRD), absorption, luminescence and emission spectra were investigated. XRD pattern was recorded using a Panalytical X'Pert pro X-ray powder diffractometer in the 2θ range of 10°–80°, with nickel-filtered Cu Kα1 radiation (λ = 1.54056 A). Rietveld refinement and bond length analysis were performed using WinPlotr and BondStr Software from the FullProf software package. Absorption spectra were measured in transmission mode by a Varian 5E UV-VIS-NIR spectrophotometer with the spectral bandwidth set to 0.5 nm. Photoluminescence excitation (PLE) and luminescence (LE) spectra and Luminescence decay (LD) curves were collected at Edinburgh Instruments FLS980 fluorescence spectrometer equipped with a 450 W xenon lamp as the excitation source.

## 3. Results and discussion

The sintering of $Cr^{4+}$:YAG ceramics occurs during vacuum sintering, with $Cr^{4+}$ ions occurring during a subsequent air annealing step. Before presenting the main results of this work, it is important to describe the synthesis process of $Cr^{4+}$:YAG ceramic and the formation mechanism of $Cr^{4+}$ ions. A characteristic feature of $Cr^{4+}$:YAG materials is the need for charge compensation additives to stabilize $Cr^{4+}$ ions. Typically, $Cr_2O_3$ is used as the chromium source. The synthesis is carried out through solid-state reaction sintering in vacuum, followed by air annealing. At present, there is no consensus in the scientific community regarding the exact mechanism of $Cr^{4+}$ ions formation. However, a generally accepted model includes the following steps:

**During vacuum sintering:**

1. *Chromium incorporation*
   - Chromium is initially incorporated into YAG lattice as $Cr^{3+}$ in octahedral sites.
     - $$Cr_2O_3 + 2Al_A^* \rightarrow 2Cr_A^* + Al_2O_3 \qquad (1)$$
2. *Charge compensation by $Ca^{2+}$ ions*

- $Ca^{2+}$ is introduced to stabilize the charge imbalance thorough the formation of $[Ca_C^{\bullet} ... V_O^{*}]$ charge neutral complex (though the exact nature of this complex remains under discussion)

$$\ll O \gg + CaO \rightarrow O_O + [Ca_Y' ... {}^1/_2 V_O^{\bullet\bullet}] \quad (2)$$

**During air annealing:**

3. *Oxygen incorporation and vacancy elimination*
   - Oxygen from the ambient atmosphere is incorporated into the lattice, eliminating the oxygen vacancies and breaking the $[Ca_C^{\bullet} ... V_O^{*}]$ charge neutral complex.

$$[Ca_Y' ... {}^1/_2 V_O^{\bullet\bullet}] + {}^1/_4 O_2 \rightarrow Ca_Y' + {}^1/_2 O_O^{*} \quad (3)$$

4. *Oxidation of $Cr^{3+}$ to $Cr^{4+}$*
   - $Ca^{2+}$ remaining from destroyed complex promotes the oxidation of $Cr^{3+}$ (in octahedral site) to $Cr^{4+}$, forming a new $[Ca_C^{\bullet} ... Cr_A^{*}]$ charge neutral complex.

$$[Ca_Y' ... {}^1/_2 V_O^{\bullet\bullet}] + {}^1/_4 O_2 + Cr_A^{*} \rightarrow [Ca_Y' ... Cr_A^{\bullet}] + {}^1/_2 O_O^{*} \quad (4)$$

5. *Cation exchange to form tetrahedral $Cr^{4+}$*
   - Finaly, the octahedrally coordinated $Cr^{4+}$ in undergoes a cation exchange with tetrahedrally coordinated $Al^{3+}$, resulting in the formation of tetrahedral $Cr^{4+}$.

$$[Ca_Y' ... Cr_A^{\bullet}] + Al_D^{*} \rightarrow [Ca_Y' ... Cr_D^{\bullet}] + Al_A^{*} \quad (5)$$

Chromium and calcium are incorporated into YAG separately during vacuum sintering and then form a charge-neutral complex during air annealing, recharging $Cr^{3+}$ into the tetravalent state. The presented model for $Cr^{4+}$ ion formation is a simplified representation and does not capture all the details. A more comprehensive explanation of $Cr^{4+}$ ion formation mechanism can be found in our earlier publications [12,13]. Based on the results previously reported for $Cr^{3+}$ ions [13], it is anticipated that the variations in charge compensation additives have only a minor effect on the spectroscopic properties of $Cr^{4+}$ ions. However, the present study does not focus on the influence of the charge compensation mechanism on the spectroscopic behavior of $Cr^{4+}$. In this work, the term Cr:YAG refers to YAG doped with $Cr^{3+}$, $Cr^{4+}$, and $Me^{2+}$ ions ($Me^{2+}$ - $Ca^{2+}$ or/and $Ca^{2+}$). The final part of the paper is dedicated to comparing the spectroscopic properties of Cr;YAG transparent ceramic and single crystal, based on literature data [14].

*3.1 Microstructure*

XRD studies confirmed the formation of pure YAG material, exhibiting disordered octahedral and tetrahedral sites. XRD pattern confirmed the presence of $Y_3Al_5O_{12}$ (Ia3d space group), with a lattice parameter 12.0242 ± 0.0007 Å. No impurity phases were found in the sample (Fig. 3). The chemical formula of YAG can be schematically written as $C_3A_2D_3O_{12}$, where C, A, and D denote dodecahedral, octahedral, and tetrahedral sites with $D_2$, $S_6$, and $S_4$ point symmetries (in Schoenflies notation), respectively [15,16]. The spectroscopic properties of $Cr^{4+}$ ions are influenced by the symmetry of the occupied sites. Deviation from the perfect octahedral or tetrahedral symmetry may arise from distortion in $Cr^{4+}$-$O^{2-}$ bond length and/or $O^{2-}$-$Cr^{4+}$-$O^{2-}$ bond angles. These structural distortions led to additional crystal field splitting, which contributes to the unique optical characteristic of $Cr^{4+}$ ions [17]. The calculated $Cr^{4+}$-$O^{2-}$ bond lengths in octahedral sites (occupied by $Al^{3+}$, $Cr^{3+}$, and $Cr^{4+}$) were found to be identical, with the average value being 1.937(4) Å. Similarly, the $Cr^{4}$-$O^{2-}$ bond lengths in tetrahedral sites (occupied by $Al^{3+}$ and $Cr^{4+}$) were also consistent, averaging 1.828(4) Å. It should be noted that the calculations were based on the following ionic radii (in Å): $Y^{3+}$ (0.930), $Al^{3+}$ (0.500), $Cr^{3+}$ (0.690), $Cr^{4+}$ (0.800), $O^{2-}$ (1.400) [18].

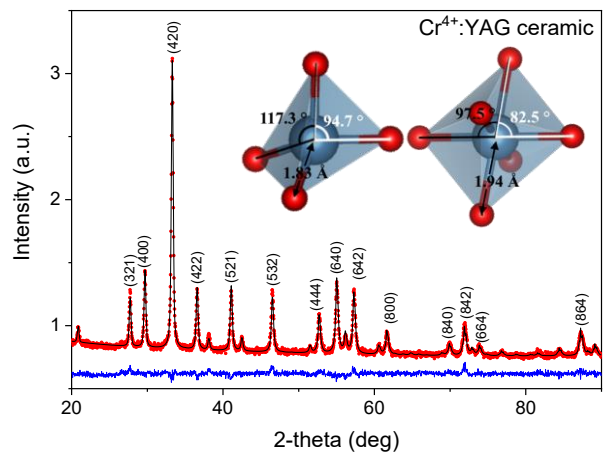

Fig. 3: XRD pattern of $Cr^{4+}$:YAG transparent ceramics (red dots), results of Rietveld refinement analysis (black curve), and the difference between experimental and the calculated pattern (blue curve).

Deviation from perfectly symmetrical sites is primarily attributed to variations in bond angles. Perfectly symmetrical octahedral and tetrahedral bond angles are 90°, and ~109.5°, respectively. Distortions introduce a variety of bond angles. Octahedral and tetrahedral sites possess fifteen and six possible combinations of $O^{2-}$-$Cr^{4+}$-$O^{2-}$ bond angle combinations, respectively. In the studied sample, the tetrahedral site exhibited four bound angles of 117.3(4)° and two bond angles of 94.7(3)°. For the octahedral sites, three types of bond angles were observed: three angles of 180.0(4)°, six angles of 82.5(3)°, and six angles of 97.5(3)°. The distortion of the tetrahedral site alters its symmetry toward that of an elongated cube. This deformation gives rise to three classes of $Cr^{4+}$ centers, oriented along the crystallographic axes [001], [010], and [100] [9].

The deviation from symmetry in tetrahedral sites leads to the splitting of $Cr^{4+}$ electron energy levels, which complicates the interpretation of their spectroscopic properties. $Cr^{4+}$ ions in tetrahedral sites of $T_d$ symmetry exhibit a $^3A_2$ ground state and $^3T_2$, $^3T_1$, and $^3T_1$ lowest excited states. Lowering the symmetry causes each of these energy levels to split into two components. A schematic illustration of this splitting is shown in Fig. 4 [19,20]. It should be noted that the energy level diagram is a simplified representation. The available literature data is insufficient to unambiguously assign all observed spectral bands to specific electronic transitions. Furthermore, the low symmetry of the tetrahedral site means that the spectroscopic behavior of $Cr^{4+}$ ions can not be adequately described using a single value of the crystal field strength parameter Dq [21]. This significantly complicates the prediction of the $Cr^{4+}$ energy level structure. Additionally, each electronic state undergoes further splitting due to spin-orbit interaction [14].

For $Cr^{4+}$ ions in tetrahedral coordination, transitions between the ground state and the first excited state are expected to be several orders of magnitude less intense than transitions to above-lying states. To better understand the measured spectroscopic properties, the energy level structure of $Cr^{4+}$ must be considered. In the free ion configuration, $Cr^{4+}$ has a ground state $^3F$ and the lowest excited states $^1D$, $^3P$, $^1G$. When placed in a crystal field of a tetrahedral site of $T_d$ symmetry, the $^3F$ energy level splits into three components, $^3F \rightarrow {}^3A_2 + {}^3T_2 + {}^3T_1$. With $^3A_2$ as the ground state, the $^3A_2 \leftrightarrow {}^3T_1$ transition should be electric dipole allowed, whereas the $^3A_2 \leftrightarrow {}^3T_2$ transition would only be magnetic dipole allowed. Consequently, the $^3A_2 \leftrightarrow {}^3T_2$ transition is expected be several orders of magnitude less intense than the $^3A_2 \leftrightarrow {}^3T_1$, transition [17]. When the site symmetry is reduced to $D_{2d}$, the ground and excited states undergo further splitting. The $^3A_2$ ground state becomes $^3B_1$, the $^3T_2$ levels split into $^3B_2$ and $^3E$, the $^3T_1$ level splits into $^3A_2$ and $^3E$.

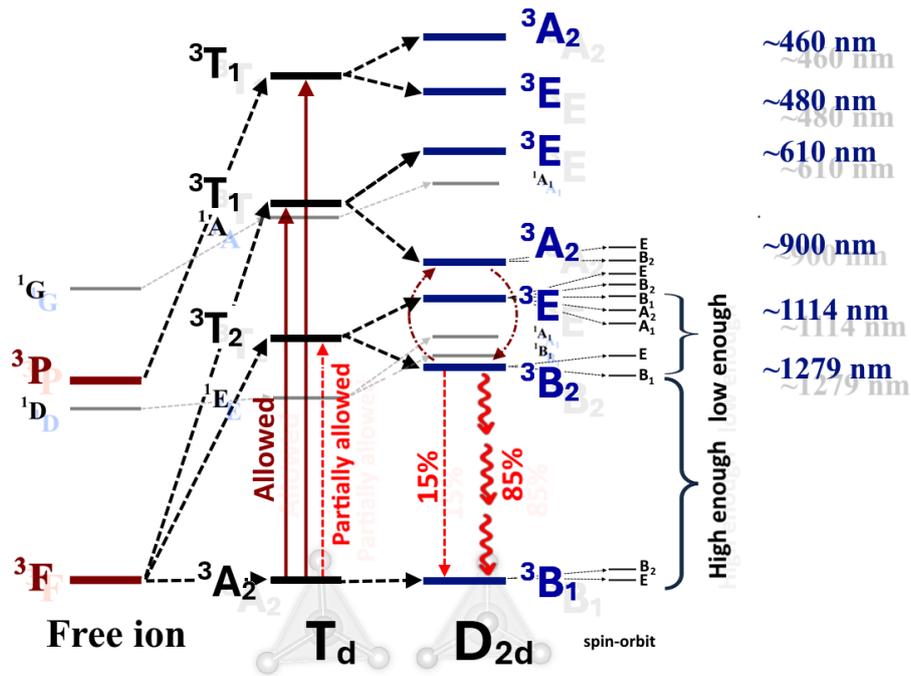

Fig. 4: Schematic illustration of the energy-level scheme of the tetrahedrally coordinated $Cr^{4+}$ ion in $T_d$ symmetry (regular tetrahedron) and $D_{2d}$-symmetry (tetragonal distorted tetrahedron), and in $D_{2d}$-symmetry, including spin-orbit splitting.

The detailed nature of $Cr^{4+}$ energy level structure may be further investigated using polarized spectroscopy, as previously demonstrated in the literature. $^3B_1(^3A_2) \rightarrow ^3E(^3T_2)$ transition at 1114 nm is electric dipole allowed in the (100) and (010) polarization direction [17]. However, the $^3B_1(^3A_2) \rightarrow ^3E(^3T_1)$ transition at 610 nm is expected to be stronger than $^3B_1(^3A_2) \rightarrow ^3E(^3T_2)$ transition at 1114 nm. This is because, in $T_d$ symmetry, the $^3A_2 \rightarrow ^3T_1$ transition was already electric dipole allowed, and $^3A_2 \rightarrow ^3T_2$ transition was only magnetic dipole allowed. The $^3B_1(^3A_2) \rightarrow ^3A_2(^3T_1)$ transition at 900 nm is electric dipole allowed in (001) polarization, while the $^3B_1(^3A_2) \rightarrow ^3B_2(^3T_2)$ (1279 nm) transition is only magnetic dipole allowed ($R_Z$). A comprehensive study on polarization-dependent spectroscopy of $Cr^{4+}$:YAG was previously reported by Eiler et. all [17]. The energy level diagram of $Cr^{4+}$ ions in the YAG lattice is shown in Fig. 4. The energy level diagram also includes two additional energy levels, $^1E$ and $^1A$, which arise from the splitting of $^1D$ and $^1G$ free-ion levels. Absorption into these states is difficult to observe due to the forbidden nature of the corresponding transitions.

*3.2 Absorption*

The absorption spectra of $Cr^{4+}$:YAG consist of contributions from both octahedral and tetrahedral $Cr^{4+}$ ions, absorbing in the visible (Vis) and near-infrared (NIR) regions, respectively. Optical absorption spectra were collected in the temperature range of 5 K – 300 K. Due to the sample thickness of 4 mm, it was difficult to collect accurate data for highly absorbing centers in the region below 300 nm. Detailed absorption spectra of $Cr^{3+}$:YAG and $Cr^{3+},Cr^{4+}$:YAG transparent ceramics are available in our earlier work [13]. The collected absorption spectra can be divided into two regions: 300–600 nm and 600–1300 nm, corresponding to the absorption of $Cr_A^{\bullet}$ ($Cr^{4+}$ ions in the octahedral site) ions and $Cr_D^{\bullet}$ ions ($Cr^{4+}$ ions in a tetrahedral site), respectively [4]. An increase in the temperature from 5 K to 300 K resulted in minimal changes in the absorption of $Cr_A^{\bullet}$ ions. The observed deviations were within a few percent and are likely attributed to changes in the refractive index of the ceramic.

Tetrahedral $Cr^{4+}$ ions are characterized by broad and narrow absorption lines; instead, octahedral $Cr^{4+}$ ions exhibit the presence of only narrow lines. The detected $Cr^{4+}$ absorption bands were assigned to the corresponding electronic transitions based on literature data [20]. However, some recorded $Cr^{4+}$ absorption bands could not be confidently assigned to specific transitions. For example, the absorption features detected at 1234 nm and 1238 nm originate from $Cr^{4+}$ ions [11], but their assignment to a particular electronic transition remains uncertain (Fig. S1). Furthermore, the proposed assignments for some transitions might differ from the actual ones. $Cr^{4+}$ ions in tetrahedral coordination of $T_d$ symmetry typically exhibit absorption bands of $^3A_2 \rightarrow {}^3T_1(^3P)$ (~470 nm), $^3A_2 \rightarrow {}^3T_1(^3F)$ (~620 nm), and $^3A_2 \rightarrow {}^3T_2(^3F)$ (~1000 nm) electronic transitions. However, the garnet's crystal structure is characterized by a tetragonally distorted tetrahedron, corresponding to $D_{2d}$ symmetry [22]. This low-symmetry crystal filed causes splitting of the $Cr^{4+}$ $^3A_2$, $^3T_1(^3P)$, $^3T_1(^3F)$, and $^3T_2(^3F)$ levels into several new energy levels: $^3B_1(^3A_2)$ (ground state), $^3B_2(^3T_2)$ (~1275 nm), $^2E(^3T_2)$ (~1114 nm), $^3A_2(^3T_1)(^3F)$ (~900 nm), $^3E(^3T_1)(F)$ (~610 nm), $^3E(^3T_1)(P)$ (~480 nm), and $^3A_2(^3T_1)(^3P)$ (~460 nm), see Fig. 5. The absorption lines corresponding to the $^3B_1(^3A_2) \rightarrow {}^3E(^3T_1)(P)$ (~480 nm) and $^3B_1(^3A_2) \rightarrow {}^3E(^3T_1)(P)$ (~460 nm) transitions are overshadowed by the strong absorption lines from $Cr_A^{\bullet}$ ions. However, this transition is noticeable in the excitation spectra.

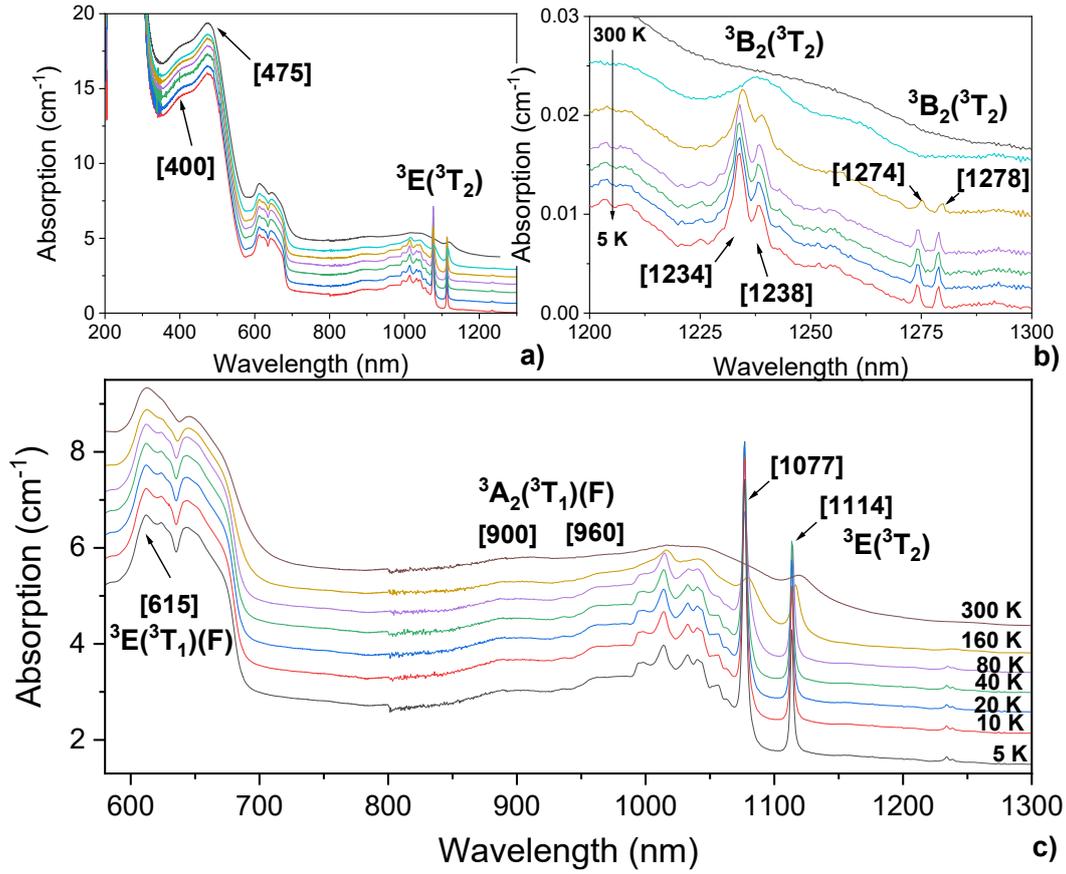

Fig. 5: Temperature dependence of $Cr^{4+}$:YAG transparent ceramic absorption spectra measured at 5-300 K in the range of λ: a) 200-1300 nm; b) 580-1300 nm; c) 1200 nm-1300 nm; The measured spectra were shifted by Δy for better visibility.

Each narrow line typically has a nearby counterpart with similar spectral parameters, which may result from crystal field splitting and/or vibronically assisted transitions. Temperature changes had minimal effects on the $Cr^{4+}$ absorption lines, with the exception of a few narrow lines. These narrow lines can be grouped based on their proximity:

- 1074 nm with 1114 nm (assigned to $^3B_1(^3A_2) \rightarrow {}^2E(^3T_2)$, separated by 300 cm$^{-1}$),
- 1234 nm with 1238 nm (assigned to $^3B_1(^3A_2) \rightarrow {}^3B_2(^3T_2)$, separated by 28 cm$^{-1}$),
- 1274 nm with 1278 nm (assigned to $^3B_1(^3A_2) \rightarrow {}^3B_2(^3T_2)$, separated by 28 cm$^{-1}$).

The double peak observed near 1235 nm could be interpreted as a 258-cm$^{-1}$ enabling mode that makes the $^3B_1(^3A_2) \leftrightarrow {}^3B_2(^3T_2)$ transition partly electric dipole allowed [17].

The temperature dependence of the lower lying absorption bands does not follow the Boltzmann law, suggesting that it does not originate from a crystal field splitting of the lower excited state. Within increasing temperature, these narrow lines – as well as the associated vibronic sidebands (particularly for the $^3B_1(^3A_2) \rightarrow {}^2E(^3T_2)$ transition) – undergo broadening. At room temperatures, these features become barely distinguishable, although their overall absorption intensity remains nearly unchanged. This suggests that these line groups are not a result of a splitting of the electronic states of $Cr_D^{\bullet}$ ions due to a lowering of crystal field symmetry or spin–orbit interaction. In such a case, their relative intensities would be expected to follow the Boltzmann law [23]. It has been previously reported that the absorption lines centered at 1077 nm and 1114 nm exhibit the same polarization dependence, indicating that both likely originate from the same transition, $^3B_1(^3A_2) \rightarrow {}^3E(^3T_2)$ [17]. A possible, though less likely, explanation is spin-orbit splitting of the $^3A_2(^3T_1)$ level, or that the line at 1077 nm represents a vibronic-assisted transition of the same electronic level [17].

*3.3 Luminescence*

Due to some similarities, the study of $Cr^{3+}$ luminescence allows for a better understanding of the $Cr^{4+}$ spectroscopic properties in $Cr^{4+}$:YAG ceramic. First, let us provide a detailed description of the spectroscopic properties of $Cr^{3+}$ ions in the YAG lattice. These properties are well-studied and can be predicted due to the extensive accumulated knowledge. In contrast, the spectroscopic behavior of $Cr^{4+}$ ions is more difficult to predict. However, both $Cr^{3+}$ and $Cr^{4+}$ exhibit a combination of broadband and narrow lines in their excitation and luminescence spectra. These similarities indicate that the approaches used to describe $Cr^{3+}$ ions can also be applied to $Cr^{4+}$. Adopting this strategy could provide deeper insight and improve the predictability of $Cr^{4+}$ spectroscopic properties.

*3.3.1 $Cr^{3+}$ luminescence*

The spectroscopic properties of $Cr^{3+}$ in the YAG lattice are characterized by broadband emission and narrow R-lines emission. Fig. 6 shows PLE and LE spectra of $Cr^{3+}$ ions measured at 5 K and 300 K. The shape of PLE spectra is determined by the spin-allowed electron transitions $^4T_{1g} - {}^4A_{2g}$ centered at 428 nm and $^4T_{2g} - {}^4A_{2g}$ centered at 588 nm, as well as the spin-forbidden transition $^2E_g - {}^4A_{2g}$ centered at 687.4 nm [24]. The transition to the $^2E_g$ level results in two weak and narrow lines, whereas transition to the $^4T_{1g}$ and $^4T_{2g}$ level generate broad excitation bands that extend from 350 nm to 650 nm [24,25]. The broad nature of the $^4T_{1g}$ and $^4T_{2g}$ bands arises from the difference in electron–lattice coupling behavior between the $^4A_{2g}$ ground state and the $^4T_{1g}/^4T_{2g}$ excited states, as described by the Huang–Rhys parameter S > 0 [6]. Due to this coupling, the configurational coordinate energy minimum (Q) of both the $^4T_{1g}$ and $^4T_{2g}$ energy parabolas is shifted relative to that of the $^4A_{2g}$ ground state. This shift leads to optical broad bands observed in both absorption and emission spectra.

Electron transitions between states occur without a change in the Q coordinate, which explains the insensitivity of the absorption line shapes to temperature variations. The transitions between two vibrational states must occur rapidly and without a significant change in the configurational coordinate Q [6]. This implies that the $^4A_{2g} \leftrightarrow {}^4T_{1g}$ and $^4A_{2g} \leftrightarrow {}^4T_{2g}$ transitions occur from the lower vibronic level of the ground state to the higher vibronic level of the excited state (Fig. S2(a)) [26,27]. As a result, the excitation bands associated with the $^4T_{1g}$ and $^4T_{2g}$ levels redshift with temperature. For example, increasing the temperature from 5 K to 300 K caused the peak excitation wavelength ($\lambda_{max}$) shift from 428 nm to 430 nm for the $^4A_{2g} \rightarrow {}^4T_{1g}$ transition and from 588 nm to 595 nm for the $^4A_{2g} \rightarrow {}^4T_{2g}$ transitions (Fig. 6(a,b)). Additionally, higher temperatures lead to broadening of both excitation bands, which can be attributed to the thermal population of the lower vibronic states within the $^4A_{2g}$ ground state. Nevertheless, the overall shape of the bands remained largely unchanged. A similar pattern was observed for $Cr^{4+}$ excitation bands, suggesting that a comparable mechanism can be applied to $Cr^{4+}$ electron transitions.

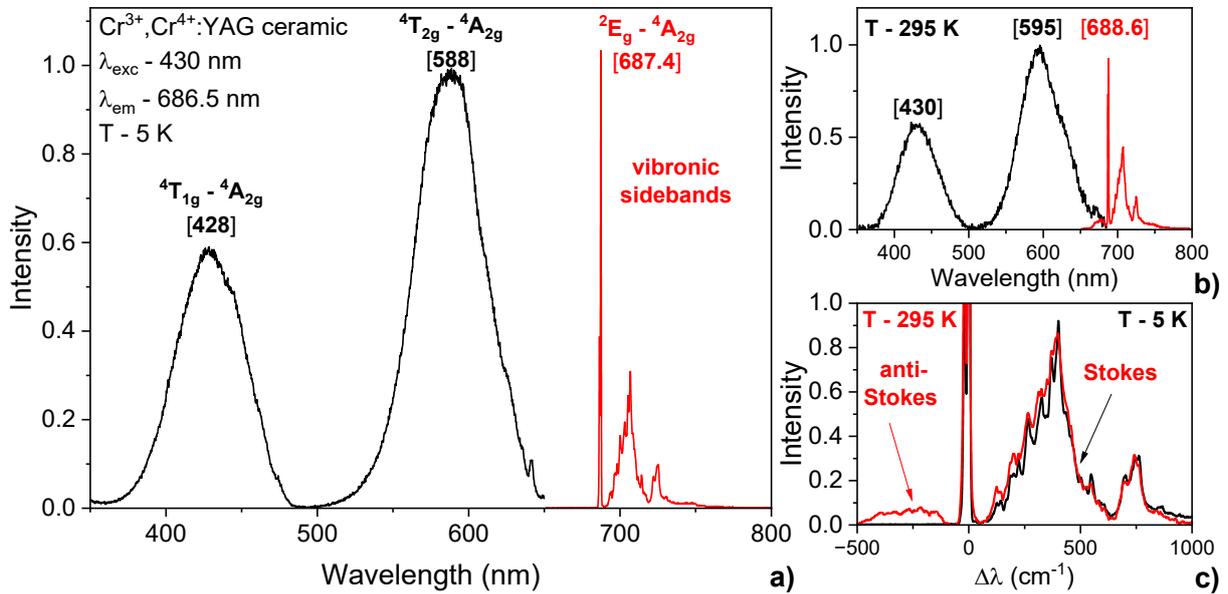

Fig. 6: PLE (black line) and LE (red line) spectra of $Cr^{3+}$ ions in the $Cr^{3+},Cr^{4+}$:YAG ceramic measured at $\lambda_{exc}$ – 430 nm; $\lambda_{em}$ – 686.5: a) T - 5 K, b) T - 295 K; c) phonon sideband of the R-line at 5 K (black line), and 295 K (red line). The x-axis values were calculated using the formula – $\Delta\lambda = \lambda(R_1\text{-line}) - \lambda$.

The similar electron–lattice coupling between the ground state and the lowest excited state results in narrow emission with a small Stokes shift, in contrast to transitions to the next higher-lying level. Due to the similar electron–lattice coupling behavior (S ≈ 0) [6], the $^4A_{2g}\leftrightarrow{}^2E_g$ transition gives narrow and sharp optical R-lines in both excitation and emission spectra [28,29], with a small Stokes shift of 4(1) cm$^{-1}$ (0.2 nm) (Fig. 7). This results in only a minor change in the R-lines emission shape along with a redshift at ~ 25 cm$^{-1}$. The R$_1$-line emission maximum shifts from 687.4 nm at 5K to 686.6 nm at 295K. The Cr$^{3+}$ broadband emission centered at ~705 nm, is overshadowed by the vibronic sidebands of the $^2E_g\rightarrow{}^4A_{2g}$ transitions, making it barely recognizable [30]. Due to the difference in the electron–lattice coupling behavior (S > 0) [6], the $^4T_{2g}\rightarrow{}^4A_{2g}$ transition begins at lower vibronic states ends at higher ones, resulting in a complex temperature-dependent pattern of Cr$^{3+}$ broadband emission parameters. Detailed analysis of these changes can be found in the literature [13].

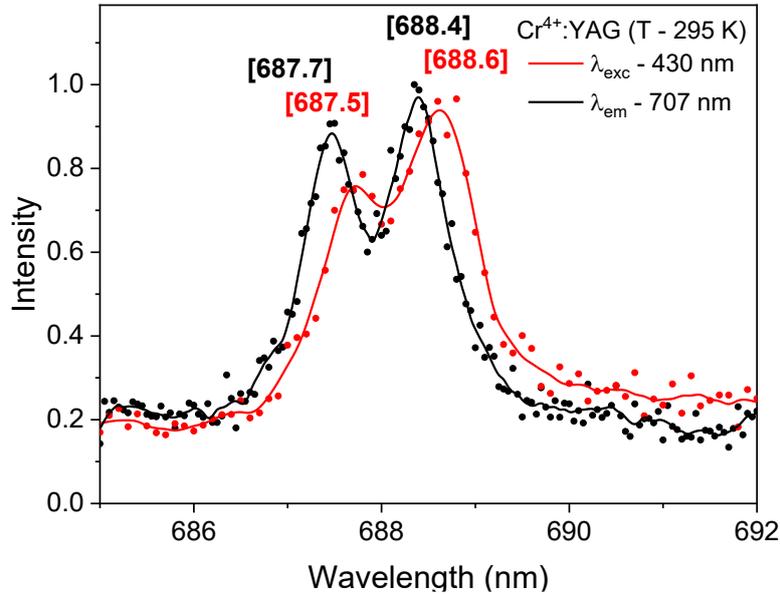

Fig. 7: PLE and LE spectra of Cr$^{3+}$ ions in Cr:YAG transparent ceramic measured at $\lambda_{em}$ – 707 nm, $\lambda_{exc}$ – 430 nm, and T – 295 K, corresponding to the $^4A_{2g}\leftrightarrow{}^2E_g$ electron transitions. Lines are included for visual guidance.

An interesting feature is that the Stokes side bands shape of the R-lines emission is nearly identical at both low and room temperatures. One characteristic feature of the $^4A_{2g}\leftrightarrow{}^2E_g$ transition is the presence of vibronic sidebands. Vibronic sidebands arise during the $^2E_g\rightarrow{}^4A_{2g}$ transition due to the absorption or emission of lattice phonons, leading to the appearance of anti-Stokes and Stokes vibronic sidebands. Fig. 6(c) shows the $^4A_{2g}\leftrightarrow{}^2E_g$ emission spectra measured at 5 K and 295 K. The x-axis values were calculated using the formula: $\Delta\lambda = \lambda(R_1\text{-line}) - \lambda$, (in cm$^{-1}$). Room temperature $^2E_g \rightarrow {}^4A_{2g}$ emission spectrum was obtained by subtracting of $^4T_{2g} \rightarrow {}^4A_{2g}$ emission band from the overall Cr$^{3+}$ emission spectrum (Fig. S2(b)). A detailed description of the Cr$^{3+}$:YAG broadband emission parameters is available in the literature [13]. A comparison of the $^2E_g \rightarrow {}^4A_{2g}$ vibronic sideband structures at 5K and 295 K reveals a similar shape for the Stokes sideband. The increase in the temperature lead to the appearance of anti-Stokes sidebands.

For Cr$^{3+}$ ions in the YAG lattice, increasing temperature leads to a stronger broadband emission contribution due to thermalization from the low-lying R-lines level. The luminescence spectra of Cr$^{3+}$ ions are determined by the positions of the $^4T_{1g}$, $^4T_{2g}$, and $^2E_g$ energy levels [31]. In the YAG lattice, Cr$^{3+}$ ions are characterized by a strong crystal field, resulting in the $^2E_g$ energy level being the lowest (Fig. S2(c)). The calculated Racah parameters are: Dq = 1680 cm$^{-1}$, Dq/B = 2.64, B = 636 cm$^{-1}$, C – 3260 cm$^{-1}$. The calculation methodology used is reported in the literature [32]. Under these conditions, only narrow spin-forbidden $^2E_g - {}^4A_{2g}$ transitions (centered at 687.4 nm) are observed at low temperature, accompanied by vibronic sidebands in the 680-750 nm range (Fig. 6(a)). With increasing temperature, a broad $^4T_{2g} \rightarrow {}^4A_{2g}$ emission appears, resulting from thermalization from the $^2E_g$ level (Fig. 6(b)). A detailed description of the temperature dependence of Cr$^{3+}$ luminescence in Cr$^{4+}$:YAG transparent ceramic can be found in the literature [13].

### *3.3.2 Cr$^{4+}$ luminescence*

The excitation spectra of Cr$^{4+}$ ions resemble they absorption spectra, confirming they role as the origin of the NIR luminescence in Cr$^{4+}$:YAG ceramic. Fig. 8 shows characteristic PLE, PL, and LD curves of Cr$^{4+}$:YAG transparent ceramic measured at T – 5 K. The PLE and PL spectra were collected using the Xe lamp, with excitation and emission slits of 15 and 10, respectively (Fig. 8(a,b)). The slits width roughly corresponds to the FWHM of the excitation/emission beam (in nm), contributing to spectral broadening. While the broad PLE spectra were largely unaffected by this broadening, PL spectra were more sensitive to these measurement conditions. PLE spectra were collected at multiple emission wavelengths, ranging from 1277 nm to 1482 nm, at a temperature of 5 K. The spectra recorded at different $\lambda_{em}$ were nearly identical to the absorption, confirming that the measured excitation features originate from $Cr_D^{4+}$ ions (Fig. 5).

The features observed in the excitation spectra match the expected patterns based on the analysis of the Cr$^{4+}$ energy level splitting in a tetrahedral environment. Cr$^{4+}$ ions located in the tetrahedral site of $T_d$ symmetry are expected to exhibit strong absorption bands corresponding to electric-dipole allowed transitions between the $^3A_2(^3F)$ ground state and the $^3T_1(^3F)$ and $^3T_1(^3P)$ excited states. In the case of a lower $D_{2d}$ symmetry, these excited states split into multiple components, which should still produce intense absorption bands. The recorded PLE spectra of $Cr^{4+}$:YAG ceramic show strong excitation bands corresponding to the crystal field component of the $^3T_1(^3F)$ level, centered at approximately 610 nm and 900 nm. In contrast, the excitation intensity of the bands attributed to the splitting of $^3T_1(^3P)$ level (located at 460 nm and 480 nm) is at least an order of magnitude lower than that of the $^3T_1(^3F)$ bands. It should be noted that the observed $Cr^{4+}$ excitation spectra partially overlap with $Cr^{3+}$ excitation bands due to the $Cr^{3+} \rightarrow Cr^{4+}$ radiative energy transfer. This further reduces the apparent intensity of the $^3T_1(^3F)$ excitation bands. Additionally, transitions between the $^3A_2$ ground state and the $^3T_1(^3P)$ level require a simultaneous change in the electronic configuration from e to $t_2$ crystal field orbitals, which significantly reduces the transition probability [14].

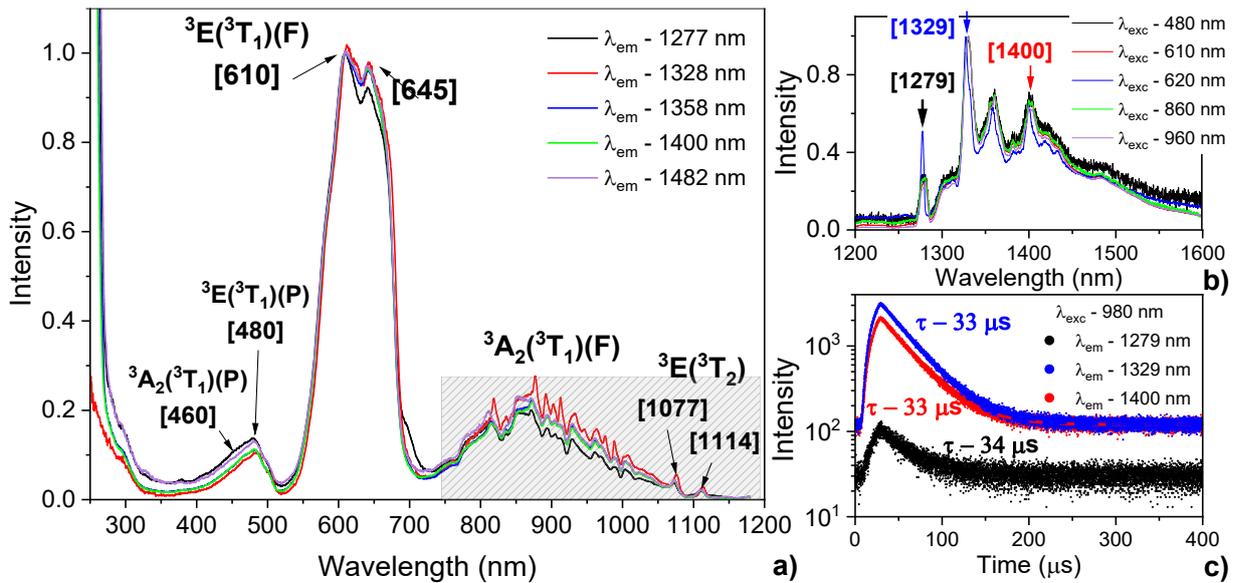

Fig. 8: Spectroscopic properties of $Cr^{4+}$:YAG transparent ceramics measured at T – 5 K: a) normalize excitation spectra measured at $\lambda_{em}$ – 1277 nm, 1328 nm, 1358 nm, 1400 nm, and 1482 nm; b) normalize emission spectra measured at $\lambda_{exc}$ – 480 nm, 610 nm, 620 nm, 860 nm, and 980 nm; c) luminescence decay curves measured at $\lambda_{exc}$ – 480 nm (diode laser), and $\lambda_{em}$ – 1278 nm, 1328 nm, and 1400 nm. The black square indicates the area with poor correction on the spectrometer sensitivity.

$Cr^{4+}$ emission spectra recorded at different excitation wavelengths are nearly identical, indicating that the emission originates solely from the lowest excited state. PLE spectra reveal the shape and position of the $^3B_1(^3A_2) \rightarrow {}^3E(^3T_1)(P)$ (~480 nm), $^3B_1(^3A_2) \rightarrow {}^3A_2(^3T_1)(P)$ (~460 nm) excitation bands, which are otherwise obscured in the absorption spectra due to the strong overlapping with $Cr_A^{4+}$ absorption. It should be noted that the recorded PLE spectra in the 700–1180 nm region do not perfectly reflect the true shape of the excitation bands of $Cr_D^{4+}$ ions, due to the low intensity of Xe lamp in this spectral range. Nevertheless, the general PLE spectral shape remains representative. PL spectra of $Cr_D^{4+}$ ions were collected at multiple wavelengths ranging from 480 nm to 960 nm, all at T – 5. No significant variations in the PL spectral shape was observed across different excitation wavelengths.

The comparable lifetimes calculated from decay curves at different emission wavelengths further indicate that the $Cr^{4+}$ luminescence originates solely from the lowest excited state. Luminescence lifetimes were measured using $\lambda_{exc}$ – 980 nm and $\lambda_{em}$ – 1278 nm, 1328 nm, and 1400 nm. It should be noted that the recorded luminescence decay curves were near the limits of the detector's sensitivity, suggesting that the actual lifetimes may differ from those calculated here. The recorded luminescence decay curves can be well fitted by a single exponential function, as described by equation (6):

$$I(t) = I_0 + A \cdot e^{(-t/\tau)} \tag{6}$$

where I(t) is the luminescence intensity at time t, $\tau$ is the decay time, and A and $I_0$ are constants. The decay time were identical for different $\lambda_{em}$ and equal to 33 µs at T – 5 K. For comparison, $Cr^{3+}$ ions might exhibit either a single exponential decay shape (due to $^2E_g$ level emission [13]) or double exponential decay (involving emission from both $^2E_g$, and $^4T_{2g}$ levels [33]). Therefore, the uniform decay time and single exponential decay shape observed here suggest that the recorded $Cr^{4+}$ emission originates only from $^3B_2(^3T_2)$ lowest excited state. This conclusion was further supported by the similarities in the excitation spectra recorded at different $\lambda_{em}$ values, as well as the consistent emission spectra across different $\lambda_{exc}$. However, the observed temperature-dependent changes in the $Cr^{4+}$ emission spectra raise questions about whether the emission truly originates from a single $^3B_2(^3T_2)$ level.

### *3.3.3 Temperature dependence*

Temperature-dependent measurements of $Cr^{4+}$ emission reveal a slight increase in intensity at low temperatures, which may suggest the presence of multiple $Cr^{4+}$ emission centers. The temperature dependence of $Cr^{4+}$ PL spectra as studied in the range of 5–600 K under laser excitation at $\lambda_{exc}$ – 980 nm (Fig, 7, Fig. S3). As the temperature increased, the emission intensity decreased, starting from approximately 40K and approaching zero at 600K. This reduction in emission intensity is attributed to an increase in the phonon-assisted nonradiative relaxation rate from the $^3B_2(^3T_2)$ lowest energy level. A small (~5%) increase in the emission intensity was observed between 5 K and 40 K. While this could be an experimental artifact, in genuine, it might suggest the presence of multiple $Cr^{4+}$ emission sites.

$Cr^{4+}$ emission features sharp and narrow zero-phonon lines (ZPLs) accompanied by vibronic sidebands, suggesting the possible presence of broadband emission from a higher-lying energy level. The emission spectra of $Cr^{4+}$ ions are primarily attributed to the electronic transitions from the $^3B_2(^3T_2)$ level. The spectra are characterized by the ZPLs centered at ~1275 nm, together with the vibronic sidebands extending up to ~1500 cm$^{-1}$. Literature reports [9–11,17,20] along with the measured PLE and lifetime data at 5K, support the conclusion that $Cr^{4+}$ luminescence spectra originate only from $^3B_2(^3T_2) \rightarrow {}^3B_1(^3A_2)$ transition. However, there remains a small possibility that the $Cr^{4+}$ luminescence includes a superposition of R-lines emission from $^3B_2(^3T_2) \rightarrow {}^3B_1(^3A_2)$ transition and broadband emission from $^3E(^3T_2) \rightarrow {}^3B_1(^3A_2)$ transition, with a peak around ~1400 nm.

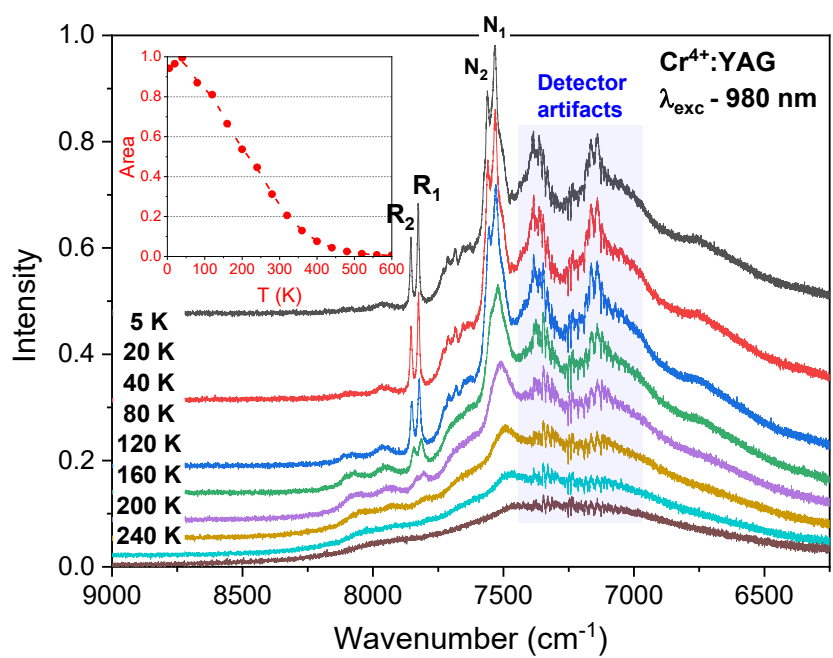

Fig. 9: Luminescence spectra of $Cr^{4+}$:YAG transparent ceramic measured at $\lambda_{exc}$ – 980 nm and temperatures from 5K to 240K. The insert shows the change in the overall $Cr^{4+}$ emission intensity.

As the temperature increases, the narrow lines broaden and eventually become indistinguishable. The collected luminesce spectra exhibits presence of two narrow emission bands along with a series of vibronic sidebands. The two narrow emission lines, centered at 1274 nm and 1278.6 nm, correspond to ZPLs associated with the $^3B_2(^3T_2)$ level. As the temperature increases, these ZPLs broaden and become indistinguishable above 200 K. A similar temperature-dependent broadening was observed for the narrow absorption lines (Fig. 5). In addition to the ZPLs, two narrow emission lines were detected at $\lambda_{max}$ – 1323.5 nm and 1328.6 nm. These are hereafter referred to as N-lines (N – neighbors). The N-lines are separated by 30 cm$^{-1}$, similarly to the separation observed for ZPLs of the $^3B_2(^3T_2)$ level. At first glance, these lines might appear to represent additional ZPLs of a different energy level. However, they also could be interpreted as a resulting from 294-cm$^{-1}$ enabling mode, that allows an otherwise magnetic dipole transition to be electric dipole allowed [17]. Another notable feature of the luminescence spectra is the vibronic sidebands accompanying the $^3B_2(^3T_2) \rightarrow ^3B_1(^3A_2)$ transition. With increasing temperature, these sidebands broaden significantly and become indistinguishable. It should be noted that the 1340–1420 nm region of the emission spectra is affected by reduced detector sensitivity, which appears as increased noise compared to the rest of the spectral range.

The redshift of Cr$^{4+}$ ZPLs with increasing temperature is approximately the same as that observed for Cr$^{3+}$ R-lines. The temperature variation influences the parameters of the $^3B_2(^3T_2) \rightarrow ^3B_1(^3A_2)$ ZPLs. These spectroscopic parameters were analyzed up to 120 K, as the ZPLs' intensity at higher temperatures was too weak for reliable analysis. An increase in the temperature from 5 K to 120 K resulted in a redshift of both R-lines by ~20 cm$^{-1}$ from 7849 cm$^{-1}$ and 7821 cm$^{-1}$ to 7833 cm$^{-1}$ and 7798 cm$^{-1}$ for ZPL$_1$ and ZPL$_2$, respectively (Fig. 10(b)). A roughly the same redshift was observed for Cr$^{3+}$ R-lines [13]. A temperature increase also caused to broadening of the ZPLs, the FWHM increased from 11 cm$^{-1}$ at 5 K to ~30 cm$^{-1}$ 120 K (Fig. 10(c)). For comparison, FWHM of Cr$^{3+}$ R-lines remained nearly constant (~30 cm$^{-1}$) over the 5–295 K.

The temperature dependence of Cr$^{4+}$ ZPLs emission intensity differs from the trend in total emission intensity, which was an unexpected result. The luminescence intensity of the Cr$^{4+}$ ZPLs decreased with rising temperature, dropping to ~30 % of the initial intensity as the temperature increased from 5 K to 120 K. However, the overall Cr$^{4+}$ luminescence intensity at 120K remained around 80% of its maximum value. Since the total Cr$^{4+}$ luminescence is a superposition of ZPLs originating from the $^3B_2(^3T_2)$ level and their vibronic sidebands, common decrease in the intensity of both component would be expected. Therefore, the observed discrepancy between the drop in ZPLs and the relatively stable overall luminescence intensity ai the 5-120 K range is unexpected. This inconsistency remains unexplained in the present study. A highly unlikely explanation is energy transfer from the $^3B_2(^3T_2)$ level to the higher-lying $^3E(^3T_2)$ state, which might be responsible for broadband emission centered around 1400 nm.

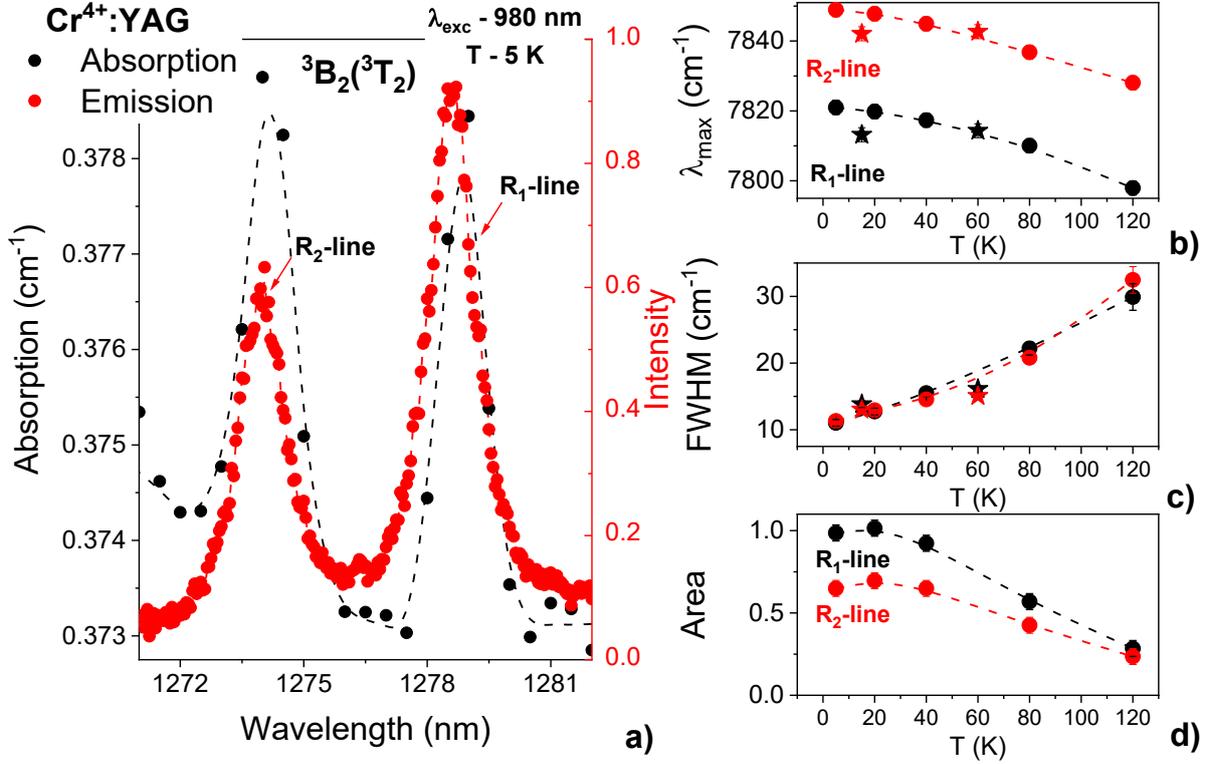

Fig. 10: (a) Absorption (black dots) and luminescence (red dots) spectra measured at $\lambda_{exc}$ – 980 nm, T – 5K. Temperature dependence of $Cr^{4+}$ - ZPLs in $Cr^{4+}$:YAG transparent ceramic (b) $\lambda_{max}$, (c) FWHM, (d) band area (stars represent the spectroscopic parameters of ZPLs for $Cr^{4+}$ ions in Cr:YAG single crystal [14]). Dashed lines are included for visual guidance.

In ceramics, $Cr^{4+}$ ZPLs exhibit a larger redshift and more pronounced brooding compared to the single crystal. The recorded $Cr^{4+}$ ZPLs parameters were compared with those from Cr:YAG single crystal [14]. In the crystal, the detected $ZPL_1$ and $ZPL_2$ emission maxima were observed at 7842 cm$^{-1}$ and 7814 cm$^{-1}$, respectively, at 15K (inset of Fig. 11). Increasing the temperature to 60 K caused no changes to the emission maxima (Fig. 10(b)). In contrast, the ceramic showed a redshift of $\lambda_{max}$ by 8 cm$^{-1}$ over the same temperature interval. The FWHM of the ZPLS in the crystal increased only 2 cm$^{-1}$ from 15 K to 60 K, while the ceramics exhibit broader $\Delta_{FWHM} \sim 6$ cm$^{-1}$ over the same $\Delta T$ (Fig. 10(d)). However, the most notable difference lies in the change in the $ZPL_1$ to $ZPL_2$ intensity ratio.

The detected ZPLs intensity ratio in the ceramic differs from the trend reported for the single crystal, raising questions about the origin of these lines. Fig. 11 shows the temperature dependence of $Cr^{4+}$ ZPLs intensity ratio for the single crystal (blue dot) and the ceramic (red dot). In the crystal, increasing the temperature from 15 K to 60 K led to a decrease in the $ZPL_1/ZPL_2$ ratio from ~2.75 to ~1.5 [14]. In contrast, the ceramic showed a smaller drop in this ratio, changing from 1.5 to 1.38 over the same temperature range. This suggests that the ceramic exhibits a smaller temperature-dependent variation of $ZPL_1/ZPL_2$ ratio compared to the single crystal. The observed discrepancies between the ceramic and the single crystal [14] may raise questions about the nature of the $Cr^{4+}$ ZPLs. It should be noted that the $Cr^{4+}$ ZPLs parameters for the single crystal were extracted from previously published graphs [14].

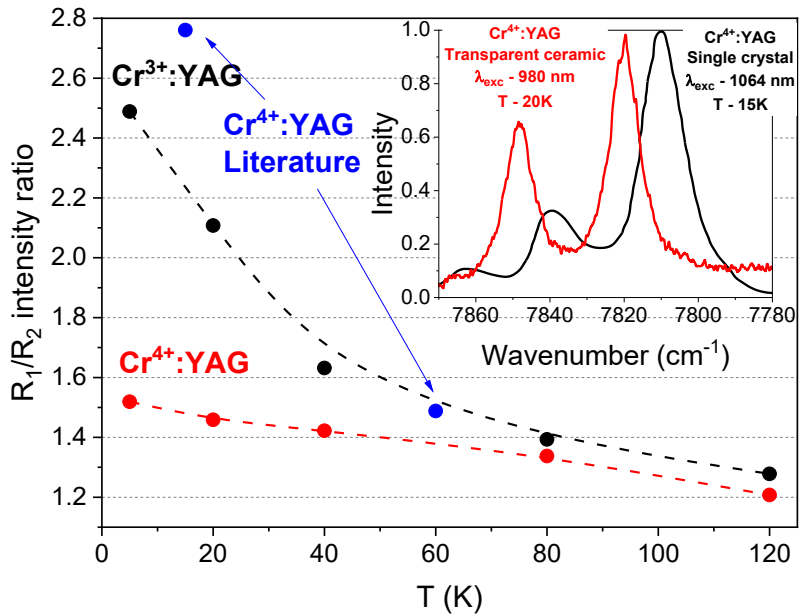

Fig. 11: $R_1/R_2$ luminescence intensity ratio of $Cr^{3+}$ (red dots), $Cr^{4+}$ (black dots) in Cr:YAG transparent ceramic, and $Cr^{4+}$ ions in Cr:YAG single crystal [14]. Insert shown $Cr^{4+}$ ZPLs emission spectra of Cr:YAG transparent ceramic (T – 20 K), and Cr:YAG single crystal (T – 15 K) [14]. Dashed lines are included for visual guidance.

The rapid temperature-dependent change in the ZPLs emission ratio observed in the single crystal suggests that these lines result from spin-orbit splitting of the lowest excited state. Kuck, et all reported the presence of two ZPLs centered at $\lambda_{max}$ - 7814 $cm^{-1}$ and 7842 $cm^{-1}$ ($\Delta\lambda$ - 28 $cm^{-1}$) in $Cr^{4+}$:YAG single crystal [14]. This splitting of the $^3B_2(^3T_2)$ absorption and emission lines was attributed to transitions into the two spin-orbit components of the excited state. In contrast, the spin-orbit splitting of the $^2B_1(^3A_2)$ ground state is expected to be much smaller, about 2 cm$^{-1}$ [14]. The observed variations of the ZPL$_1$/ZPL$_2$ intensity ratio for the crystal (blue dots in Fig. 11) support the interpretation that the ZPLs originate from the spin-orbit splitting of the lowest excited $^3B_2(^3T_2)$ level. The rapid increase in the relative intensity of the ZPL$_2$ with increasing temperature was attributed to the thermally driven population of the higher lying ZPL$_2$ component, as governed by Boltzmann statistics [34].

A similar phenomenon has been observed for Cr$^{3+}$ ions in the YAG lattice, where increasing temperature alters the R-lines emission ratio due to thermal coupling of two spin-orbit components. The typical temperature-dependent behavior of the R$_1$/R$_2$ intensity ratio for the Cr$^{3+}$ ions (black dots in Fig. 11) follows a Boltzmann distribution. Let's consider the influence of the temperature on the Cr$^{3+}$ R-liens emission behavior. Imperfections in the octahedral sites of the YAG lattice result in the splitting of the Cr$^{3+}$ energy levels into sublevels. Specifically, the $^4A_{2g}$ ground state is split at ~0.4 cm$^{-1}$, while the lowest excited state, $^2E_g$ split at ~38 cm$^{-1}$. Electrons in the $^2E_g$ level might occupy either the lower or the higher sublevel. Emission from the lower sublevel results in the R$_1$-line, while the emission from the higher sublevel results in R$_2$-line. At low temperature, the lower sublevel is more populated, leading to stronger R$_1$-line emission. As the temperature increases, the population of the higher sublevel increases, as governed by Boltzmann statistics, resulting in an increased R$_2$-line emission intensity until thermal equilibrium is reached [34]. Eventually, thermal quenching leads to the decline in the intensities of both R-lines beyond a critical temperature.

The changes in ZPLs ratio in the ceramic differ from those in the single crystal, suggesting that the two ZPLs may not originate from spin-orbit splitting of the lower existed state. Temperature-dependent changes in the Cr$^{4+}$ ions ZPL$_1$/ZPL$_2$ ratio observed in the ceramic sample differ from those of Cr$^{3+}$ ions in the ceramic and from Cr$^{4+}$ ions in the single crystal. While a reduction of the Cr$^{4+}$ ZPL$_1$/ZPL$_2$ emission ratio is observed, the initial ratio is significantly lower than expected. Moreover, both ZPL$_1$ and ZPL$_2$ emission intensities follow a similar temperature-dependent pattern (Fig. 10(d)). This behavior suggests that the observed emission lines at 7849 cm$^{-1}$ and 7821 cm$^{-1}$ may originate from distinct Cr$^{4+}$ ions, rather than from the spin-orbit splitting of the same $^3B_2(^3T_2)$ and/or $^3B_1(^3A_2)$ levels. A possible explanation lies in the presence of differently oriented Cr$^{4+}$ optical centers aligned along crystallographic axes [001], [010], and [100] [10]. Although, spin-orbit splitting of the $^3B_2(^3T_2)$ level, as previously proposed for the single crystal [14], can not be entirely excluded, the alternative involving oriented centers merits consideration.

The observed absorption and emission features of $Cr^{4+}$ ions may be explained by the presence of three classes of $Cr^{4+}$ ions. The optical characteristics of $Cr^{4+}$:YAG materials are influenced by the distortion of the tetrahedral site. In a perfect tetrahedral $T_d$ crystal field, $^3F$ energy level of free $Cr^{4+}$ ions splits into $^3A_2$, $^3T_2$, and $^3T_1$ energy levels. When the symmetry is lowered to $D_{2d}$, further splitting occurs: the ground state $^3A_2$ become $^3B_1(^3A_2)$, while the lowest excited state $^3T_2$ splits into $^3B_2(^3T_2)$ and $^3E(^3T_2)$ components [17]. This distortion is associated with the elongated cubic symmetry of the site, leading to formation of three classes of $Cr^{4+}$ ions, each aligned along a different crystallographic axis. Pressure-dependent absorption and emission studies on a single crystal confirmed the presence of such centers [17]. Eilers et. all demonstrated the difference in the emission spectra using polarized light propagating along one and polarized along another crystallographic axis, and separation of $^3E(^3T_2)$ energy level components under applied pressure [17]. However, due to the general isotropy of $Cr^{4+}$:YAG optical properties, clear experimental evidence confirming that the detected doublet line originate from different $Cr^{4+}$ oriented centers remains unclear.

Another possible explanation involves different types of $(CrO_4)^{6-}$ optical active centers. Our earlier studies revealed two types of $(CrO_6)^{9-}$ centers in Cr,Ca:YAG ceramics [13]. The addition of $Ca^{2+}$ ions leads to the appearance of two distinct $Cr^{3+}$ luminescence centers, one exhibiting a redshift of 27 cm$^{-1}$ [13]. In the YAG lattice, $Cr^{3+}$ ions occupies octahedral site surrounded by four dodecahedral $Y^{3+}$ ions, each at a distance of 3.36 Å. Substitution of one $Y^{3+}$ ion by $Ca^{2+}$ likely caused the observed spectral shift. A similar mechanism might be proposed for $(CrO_4)^{6-}$ centers, where $Cr^{4+}$ in a tetrahedral site is surrounded by six dodecahedral $Y^{3+}$ ions – two at 3.01 Å and four at 3.86 Å. Replacement of nearby $Y^{3+}$ ions by $Ca^{2+}$ could potentially result in different $Cr^{4+}$ centers and contribute to the observed spectral splitting. Variation in charge-compensating additives ($Ca^{2+}$ and/or $Mg^{2+}$) may also influence optical properties. However, this explanation remains speculative.

In summary, three possible mechanisms for the observed splitting of the $^3B_2(^3T_2)$ energy level in $Cr^{4+}$ have been proposed.

- Spin-orbit splitting of the $^3B_2(^3T_2)$ level into two component [14].
- Presence of differently oriented $Cr^{4+}$ centers alignment with distinct crystallographic axes [17].
- Formation of different $(CrO_4)^{6-}$ optical centers due to local distortion from charge-compensating additives [13].

At present, the nature of the $^3B_2(^3T_2)$ energy level splitting in $Cr^{4+}$ ions remains unresolved. A better understanding of these luminescence properties is crucial for improving of both passive Q-switched and continuous wave (CW) lasers based on $Cr^{4+}$:YAG materials.

*3.4 Laser performance*

$Cr^{4+}$:YAG transparent ceramic used for passive Q-switched lasers. The operation principle of passive Q-switched lasers is based on the use of optical (phototropic) materials that change their transparency according to the absorbed energy. In the unexcited state, they have a low transmission coefficient for radiation at the laser operation wavelength, that is, they causes large losses. They absorb the radiation emitted by laser ions, preventing the strong photon flux inside the cavity. This increases the generation threshold and therefore allows the accumulation of the pumping energy by laser ions. After reaching the generation threshold strong photon flux starts to build up in the resonator, which causes saturation of absorption, so they become transparent in a short time. This rapidly reduces the losses in the cavity and accumulated energy released in a giant pulse, after which the phototropic absorber begins recovery for the next cycle [5].

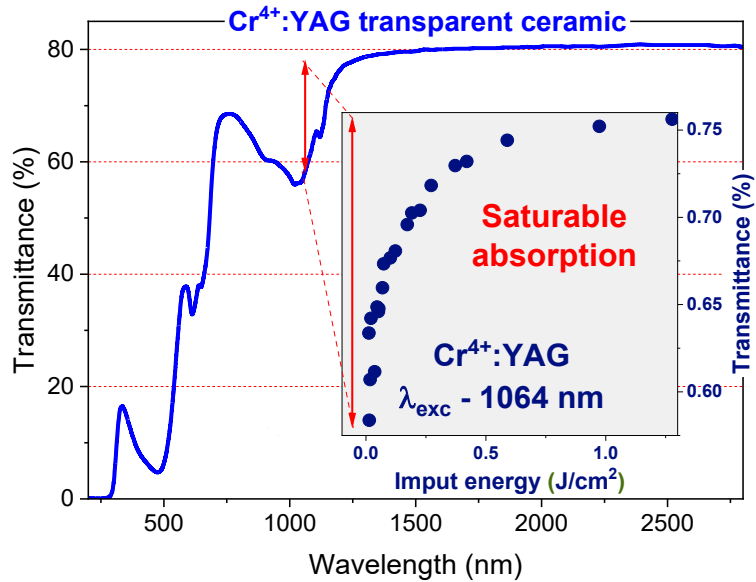

Fig. 12: Transmittance spectra of the measured $Cr^{4+}$:YAG transparent ceramic. The insert shows the change in transmittance of $Cr^{4+}$:YAG transparent ceramic under irradiation by 1064 nm pulsed laser (red circles) as a function of the input energy fluence [4].

The laser properties of passive Q-switched lasers depend on many parameters, such as the properties of the reflection mirrors in the resonator, spectral range, pumping power, etc. The most important parameter among the others is the difference in the absorption initially and after saturation of the phototropic material [8]. Simplified, the higher this difference, the higher energy per pulse can be obtained. Fig. 12 shows the change in $Cr^{4+}$ absorption with the absorbing power [4]. The change in the radiation energy of pulsed 1064 nm $Nd^{3+}$:YAG laser caused to change in the absorption of this light by $Cr^{4+}$ ions. For the specific sample, the increase in the input energy fluence to 1.3 J/cm$^2$ caused to increase in transparency of $Cr^{4+}$:YAG ceramic from 56 % to 76% due to the saturation of $Cr^{4+}$ absorption.

The laser performance of $Cr^{4+}$:YAG/$Nd^{3+}$:YAG passive q-switched laser is determined by a lot of parameters, including optical resonator length, mirrors' reflectivity, pump power, $Nd^{3+}$ ions concentrations, $Nd^{3+}$ ions lifetimes, $Cr^{4+}$ ions lifetime, $Cr^{4+}$ ions concentration, $Nd^{3+}$:YAG length, $Cr^{4+}$:YAG length, etc [35,36]. The most important parameter among the other is the ability $Cr^{4+}$ ions to suppress $Nd^{3+}$:YAG laser generation by absorbing their emission. Fig. 13 shows the time profile of laser pulse energy of $Nd^{3+}$:YAG/$Cr^{4+}$:YAG Q-switched laser with $Cr^{4+}$ absorption of 1 OD (Fig. 13(a)) and $Cr^{4+}$ absorption of 0.6 OD (Fig. 13(b). If the $Cr^{4+}$ ions' absorption is high enough to suppress $Nd^{3+}$:YAG continuous wave generation, the mono impulses are generated (Fig. 13(a). If the absorption of $Cr^{4+}$ ions is not enough, CW generation will be detected with the oscillation in the power generation (Fig. 13(b)).

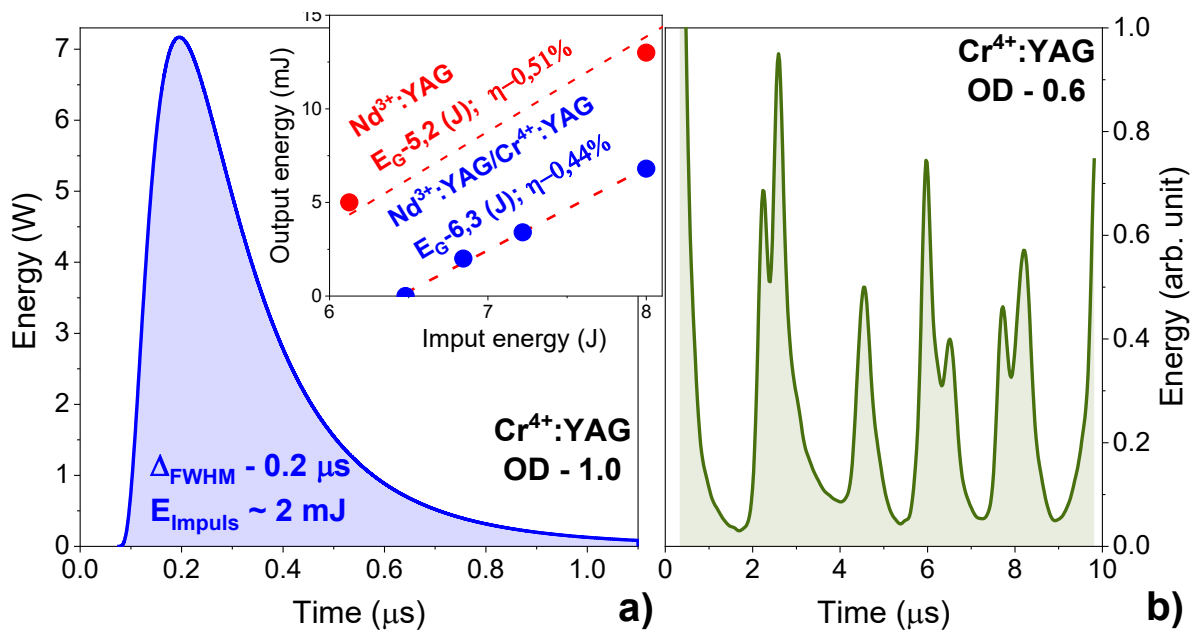

Fig. 13. Time profile of laser pulse energy of $Nd^{3+}$:YAG/$Cr^{4+}$:YAG Q-switched laser with a) $Cr^{4+}$ absorption of 1 OD, b) $Cr^{4+}$ absorption of 0.6 OD. Insert the show-lasing performance of $Nd^{3+}$:YAG active elements using the scheme with lamp pumping in free-lasing mode (red dots) and passive Q-switching mode using $Cr^{4+}$:YAG saturable absorber.

The laser properties were measured by placing $Cr^{4+}$:YAG ceramics inside the resonator. With only $Nd^{3+}$:YAG single crystal road, the laser efficiency was η - 0.51% and power generation threshold $E_g$ – 5.2 J. Placement $Cr^{4+}$:YAG ceramics increases the laser generation threshold $E_g$ to 6.3 J and decreases the laser efficiency to η - 0.44%. The peak power generation was 7 W. It should be noted that the used single crystal with $Cr^{4+}$ ions absorption at 1064 nm of 1.3 OD and without AR coating demonstrates the pulse energy of 30 mJ. At the current moment, $Cr^{4+}$:YAG transparent ceramics are far from perfect. To improve their laser performance, further investigations are required.

## 5. Conclusion

The spectroscopic properties of $Cr^{4+}$:YAG transparent ceramics were investigated in details. XRD analysis confirmed the phase purity of the sample and revealed the presence of a disordered tetrahedral site. The absorption spectra displayed characteristic features of $Cr^{4+}$ ions occupying both octahedral sites (absorption bands at 300-600 nm) and tetrahedral sites (600-1300 nm). A notable feature of the low-temperature absorption spectra is the presence of several narrow lines centered at 1074 nm, 1114 nm, 1234 nm, 1238 nm, 1274 nm, and 1278 nm. As the temperature increases, these narrow lines – as well as their associated vibronic sidebands – undergo broadening.

Excitation spectra were nearly identical to the absorption spectra, confirming that the observed features originate from $Cr^{4+}$ ions. Emission spectra of $Cr^{4+}$ ions collected at excitation wavelengths from 480 nm to 960 nm were nearly identical. With increasing temperature, the emission intensity began to decrease starting from ~40 K and approached zero at 600 K. The spectra are characterized by the sharp and narrow zero-phonon lines centered at ~1275 nm, accompanied by vibronic sidebands extending up to ~1500 $cm^{-1}$. Photoluminescence excitation and lifetime measurements confirm that $Cr^{4+}$ luminescence originates exclusively from the $^3B_2(^3T_2) \rightarrow {}^3B_1(^3A_2)$ transition.

Two narrow emission lines centered at 1274 nm and 1278.6 nm were detected in the luminescence spectra. Temperature-dependent changes in the $Cr^{4+}$ ions $ZPL_1/ZPL_2$ ratio observed in the ceramic differ from the single crystal and those of $Cr^{3+}$ ions. Although a reduction of the $ZPL_1/ZPL_2$ ratio is observed, the initial ratio is significantly lower than expected. Additionally, both $ZPL_1$ and $ZPL_2$ exhibit similar temperature-dependent intensity profiles. These findings suggest that the two ZPLs may originate from distinct $Cr^{4+}$ ions rather than from the spin-orbit splitting of the same $^3B_2(^3T_2)$ and/or $^3B_1(^3A_2)$ levels. One possible explanation involves the presence of differently oriented $Cr^{4+}$ optical centers aligned along [001], [010], and [100] crystallographic axes. However, other explanations – such as different local environments or charge compensation mechanisms – may also contribute to the observed behavior.

**Acknowledgements**

The authors are indebted to dr. M. Blees (CoorsTek, Netherlands) for making and providing the $Cr^{4+}$:YAG samples, and also, for fruitful discussions.

**Funding**

This work was supported by Polish National Science center, grant: OPUS 23, UMO-2022/45/B/ST5/01487. This work has been co-financed by the European Union under the HORIZON.1.2 – Marie Skłodowska-Curie Actions (MSCA), topic HORIZON-MSCA-2023-SE-01 – MSCA Staff Exchanges 2023, "$Cr^{4+}$:YAG/Polymer nanocomposite as alternative materials for Q-switched lasers: properties, modeling, and applications – ALTER-Q" - Project number 101182995.

**Data availability**

The datasets used during the current study are available by the link 10.5281/zenodo.15790036 or from the corresponding author on request.

**References**

[1] P. Gluchowski, D. Hreniak, W. Lojkowski, W. Strek, Optical Properties of Cr(III) doped YAG Nanoceramics, ECS Trans 25 (2009) 113–119. https://doi.org/10.1149/1.3211168/XML.

[2] P. Gluchowski, W. Strek, Luminescence and excitation spectra of Cr3+:MgAl 2O4 nanoceramics, Mater Chem Phys 140 (2013) 222–227. https://doi.org/10.1016/j.matchemphys.2013.03.025.

[3] M. Chaika, Advancements and challenges in sintering of $Cr^{4+}$:YAG: A review, J Eur Ceram Soc 44 (2024) 7432–7450. https://doi.org/10.1016/J.JEURCERAMSOC.2024.05.050.

[4] M. Chaika, R. Lisiecki, K. Lesniewska-Matys, O.M. Vovk, A new approach for measurement of Cr4+ concentration in Cr4+:YAG transparent materials: Some conceptual difficulties and possible solutions, Opt Mater (Amst) 126 (2022) 112126. https://doi.org/10.1016/j.optmat.2022.112126.

[5]     W. Koechner, Solid-state laser engineering, Springer Nature, 2013.

[6]     J.G. Solé, L.E. Bausá, D. Jaque, An Introduction to the Optical Spectroscopy of Inorganic Solids, John Wiley and Sons, 2005. https://doi.org/10.1002/0470016043.

[7]     W. Koechner, M. Bass, Solid-State Lasers, (2003). https://doi.org/10.1007/B97423.

[8]     A. Agnesi, S. Dell'Acqua, C. Morello, G. Piccinno, G.C. Reali, Z. Sun, Diode-pumped neodymium lasers repetitively Q-switched by Cr4+:YAG solid-state saturable absorbers, IEEE Journal on Selected Topics in Quantum Electronics 3 (1997) 45–52. https://doi.org/10.1109/2944.585813.

[9]     A. Okhrimchuk, A. Shestakov, Absorption saturation mechanism for YAG:Cr4+ crystals, Phys Rev B 61 (2000) 988. https://doi.org/10.1103/PhysRevB.61.988.

[10]    V. Kartazaev, R.R. Alfano, Polarization influence of excited state absorption on the performance of Cr4+:YAG laser, Opt Commun 242 (2004) 605–611. https://doi.org/10.1016/J.OPTCOM.2004.09.007.

[11]    K.R. Hoffman, U. Hömmerich, S.M. Jacobsen, W.M. Yen, On the emission and excitation spectrum of the NIR laser center in Cr:YAG, J Lumin 52 (1992) 277–279. https://doi.org/10.1016/0022-2313(92)90031-4.

[12]    M.A. Chaika, R. Tomala, W. Strek, D. Hreniak, P. Dluzewski, K. Morawiec, P. V. Mateychenko, A.G. Fedorov, A.G. Doroshenko, S. V. Parkhomenko, K. Lesniewska-Matys, D. Podniesinski, A. Kozłowska, G. Mancardi, O.M. Vovk, Kinetics of Cr3+ to Cr4+ ion valence transformations and intra-lattice cation exchange of Cr4+ in Cr,Ca:YAG ceramics used as laser gain and passive Q-switching media, Journal of Chemical Physics 151 (2019) 134708. https://doi.org/10.1063/1.5118321.

[13]    M. Chaika, K. Elzbieciak-Piecka, O. Vovk, L. Marciniak, New explanation for oxidation-induced Cr4+ formation in garnets, Optical Materials: X 23 (2024) 100342. https://doi.org/10.1016/J.OMX.2024.100342.

[14]    S. Kück, K. Petermann, U. Pohlmann, G. Huber, Electronic and vibronic transitions of the Cr4+-doped garnets Lu3Al5O12, Y3Al5O12, Y3Ga5O12 and Gd3Ga5O12, J Lumin 68 (1996) 1–14. https://doi.org/10.1016/0022-2313(95)00088-7.

[15]    R. Tomala, K. Grzeszkiewicz, D. Hreniak, W. Strek, Downconversion process in Yb3+doped GdAG nanocrystals, J Lumin 193 (2018) 70–72. https://doi.org/10.1016/j.jlumin.2017.06.038.

[16] R. Tomala, L. Marciniak, J. Li, Y. Pan, K. Lenczewska, W. Strek, D. Hreniak, Comprehensive study of photoluminescence and cathodoluminescence of YAG:Eu3+ nano- and microceramics, Opt Mater (Amst) 50 (2015) 59–64. https://doi.org/10.1016/J.OPTMAT.2015.06.042.

[17] H. Eilers, U. Hömmerich, S.M. Jacobsen, W.M. Yen, K.R. Hoffman, W. Jia, Spectroscopy and dynamics of Cr4+:Y3Al5O12, Phys Rev B 49 (1994) 15505. https://doi.org/10.1103/PhysRevB.49.15505.

[18] I.D. Brown, D. Altermatt, Bond-valence parameters obtained from a systematic analysis of the Inorganic Crystal Structure Database, Acta Crystallographica Section B 41 (1985) 244–247. https://doi.org/10.1107/S0108768185002063.

[19] S. Kück, K. Petermann, U. Pohlmann, G. Huber, Near-infrared emission of Cr4+-doped garnets: Lifetimes, quantum efficiencies, and emission cross sections, Phys Rev B 51 (1995) 17323. https://doi.org/10.1103/PhysRevB.51.17323.

[20] A.G. Okhrimchuk, A. V. Shestakov, Performance of YAG: Cr4+ laser crystal, Opt Mater (Amst) 3 (1994) 1–13. https://doi.org/10.1016/0925-3467(94)90023-X.

[21] E. König, S. Kremer, Ligand Field, Ligand Field (1977). https://doi.org/10.1007/978-1-4757-1529-3.

[22] N.L. Borodin, A.G. Okhrimchuk, A. V. Shestakov, Polarizing Spectroscopy of Y3Al5O12, SrAl2O4, CaAl2O4 Crystals Containing Cr4+, Advanced Solid State Lasers (1992), Paper CL11 (1992) CL11. https://doi.org/10.1364/ASSL.1992.CL11.

[23] M.Y. Sharonov, A.B. Bykov, V. Petričević, R.R. Alfano, Cr4+-doped Li2CaSiO4 crystal: growth and spectroscopic properties, Opt Commun 231 (2004) 273–280. https://doi.org/10.1016/J.OPTCOM.2003.12.044.

[24] V.B. Mykhaylyk, H. Kraus, Y. Zhydachevskyy, V. Tsiumra, A. Luchechko, A. Wagner, A. Suchocki, Multimodal Non-Contact Luminescence Thermometry with Cr-Doped Oxides, Sensors 2020, Vol. 20, Page 5259 20 (2020) 5259. https://doi.org/10.3390/S20185259.

[25] P. Głuchowski, R. Pazik, D. Hreniak, W. Strek, Luminescence studies of Cr3+ doped MgAl2O4 nanocrystalline powders, Chem Phys 358 (2009) 52–56. https://doi.org/10.1016/j.chemphys.2008.12.018.

[26] A. Luchechko, V. Vasyltsiv, V. Stasiv, M. Kushlyk, L. Kostyk, D. Włodarczyk, Y. Zhydachevskyy, Luminescence spectroscopy of Cr3+ ions in bulk single crystalline β-Ga2O3-In2O3 solid solutions, Opt Mater (Amst) 151 (2024) 115323. https://doi.org/10.1016/J.OPTMAT.2024.115323.

[27] P. Głuchowski, R. Pazik, D. Hreniak, W. Strek, Luminescence properties of Cr3+:Y3Al5O12 nanocrystals, J Lumin 129 (2009) 548–553. https://doi.org/10.1016/j.jlumin.2008.12.012.

[28] V.B. Mykhaylyk, H. Kraus, L.I. Bulyk, I. Lutsyuk, V. Hreb, L. Vasylechko, Y. Zhydachevskyy, A. Wagner, A. Suchocki, Al2O3 co-doped with Cr3+ and Mn4+, a dual-emitter probe for multimodal non-contact luminescence thermometry, Dalton Transactions 50 (2021) 14820–14831. https://doi.org/10.1039/D1DT02836G.

[29] V. Stadnik, V. Hreb, A. Luchechko, Y. Zhydachevskyy, A. Suchocki, L. Vasylechko, Sol-Gel Combustion Synthesis, Crystal Structure and Luminescence of Cr3+ and Mn4+ Ions in Nanocrystalline SrAl4O7, Inorganics 2021, Vol. 9, Page 89 9 (2021) 89. https://doi.org/10.3390/INORGANICS9120089.

[30] Y. Zhydachevskyy, V. Mykhaylyk, V. Stasiv, L.I. Bulyk, V. Hreb, I. Lutsyuk, A. Luchechko, S. Hayama, L. Vasylechko, A. Suchocki, Chemical Tuning, Pressure, and Temperature Behavior of Mn4+Photoluminescence in Ga2O3-Al2O3Alloys, Inorg Chem 61 (2022) 18135–18146. https://doi.org/10.1021/ACS.INORGCHEM.2C02807/ASSET/IMAGES/LARGE/IC2C02807_0015.JPEG.

[31] L. Vasylechko, V. Stadnik, V. Hreb, Y. Zhydachevskyy, A. Luchechko, V. Mykhaylyk, H. Kraus, A. Suchocki, Synthesis, Crystal Structure and Photoluminescent Properties of Red-Emitting CaAl4O7:Cr3+ Nanocrystalline Phosphor, Inorganics (Basel) 11 (2023) 205. https://doi.org/10.3390/INORGANICS11050205/S1.

[32] Z. Song, P.A. Tanner, Q. Liu, Host Dependency of Boundary between Strong and Weak Crystal Field Strength of Cr3+ Luminescence, Journal of Physical Chemistry Letters 15 (2024) 2319–2324. https://doi.org/10.1021/ACS.JPCLETT.4C00008/ASSET/IMAGES/LARGE/JZ4C00008_0004.JPEG.

[33] P. Gluchowski, M. Chaika, Crystal-Field Strength Variations and Energy Transfer in $Cr^{3+}$-Doped GGG Transparent Nanoceramics, Journal of Physical Chemistry C 128

(2024) 9641–9651. https://doi.org/10.1021/ACS.JPCC.4C01658/ASSET/IMAGES/LARGE/JP4C01658_0011.JPEG.

[34] L. Vasylechko, V. Stadnik, V. Hreb, Y. Zhydachevskyy, A. Luchechko, V. Mykhaylyk, H. Kraus, A. Suchocki, Synthesis, Crystal Structure and Photoluminescent Properties of Red-Emitting CaAl4O7:Cr3+ Nanocrystalline Phosphor, Inorganics (Basel) 11 (2023) 205. https://doi.org/10.3390/INORGANICS11050205/S1.

[35] Моделювання та оптимізація твердотільних мікрочипових лазерів | Видавництво Львівської політехніки, (n.d.). https://vlp.com.ua/node/10763 (accessed June 12, 2025).

[36] O. Buryy, S. Ubizskii, Modeling and optimization of the YAG:Yb microchip laser passively Q-switched by YAG:Cr absorber, Optica Applicata Vol. 44 (2014) 621–636. https://doi.org/10.5277/OA140412.